\documentstyle[pra,aps,twocolumn]{revtex}
\input epsf
\begin{document}

\title{Modulational instability of bright solitary waves in
incoherently coupled nonlinear Schr\"odinger equations}

\author{Dmitry V. Skryabin
\footnote{URL: http://cnqo.phys.strath.ac.uk/$\sim$dmitry},
and William J. Firth}
\address{Department of
Physics and Applied Physics, John Anderson Building,\\ University
of Strathclyde, 107 Rottenrow, Glasgow, G4 0NG, Scotland}

\date{October 5, 1998}

\maketitle

\begin{abstract}
We present a detailed analysis of the modulational instability (MI) of
ground-state bright solitary solutions of  two incoherently
coupled nonlinear Schr\"odinger equations. Varying the relative strength of
cross-phase and self-phase effects we show existence and origin of
four branches of MI of the two-wave solitary solutions. We give a physical
interpretation of our results in terms of the group velocity dispersion (GVD)
induced polarization dynamics of spatial solitary waves.
In particular, we show that in media with normal
GVD spatial symmetry breaking changes to  polarization symmetry breaking when
the relative strength of the cross-phase  modulation exceeds a certain threshold value.
The analytical and numerical stability analyses are fully supported by an extensive
series of numerical simulations of the full model.

\end{abstract}


\section{INTRODUCTION}

The phenomenon of modulational instability (MI) can be defined as
self-induced break up of an initially homogeneous wave during its evolution
in a nonlinear medium. Study of this phenomenon has been initiated
in the 60's, when MI was predicted
in plasma physics \cite{Askaryan62}, nonlinear optics
\cite{Bespalov66} and physics of fluids \cite{Benjamin67},
and also observed experimentally
in form of filamentation of an optical beam propagating in an organic liquid
\cite{Pilipetskii65}.
Since that time MI has remained as one of the major topics
of theoretical and experimental  research in nonlinear physics
and, in particular, in nonlinear physics of conservative systems
\cite{comment1,PhysRep,Rasmussen,Berge,Akhmedbook,KivsharPHYSREP}.
 We will deal below with one classical example of such systems, unifying
 a number of previous results, and presenting new MI phenomena.
 Our approach stresses the central role of symmetries.

A general formulation of the problem of nonlinear wave propagation
via fundamental sets of equations, as for example, the  Maxwell or
Navier Stokes equations, is a very demanding task
even for modern computers. Therefore a number of simplified models
have been introduced which approximately describe either
propagation of the wave itself, e.g., KdV equation \cite{book},
or propagation of slowly varying  wave envelope,
e.g., the  Nonlinear Schr\"odinger (NLS) equation \cite{book}.

The simplest solutions of  envelope equations are continuous wave
(CW) solutions homogeneous in space and time.
CWs in single NLS equation exhibit MI in cases
when nonlinearity and group velocity dispersion (GVD) or diffraction
act in opposition, e.g. when nonlinearity is positive GVD must be
anomalous and if nonlinearity is negative GVD must be normal.
This rule changes when, accounting for   polarization, for
different directions of wave vectors, or for  different carrier
frequencies of the interacting waves,  one replaces the single NLS
by the set of incoherently coupled NLS equations. Then, if nonlinear coupling
is strong enough, MI of CWs becomes possible for any signs of
nonlinearity and GVD
\cite{Berkhoer70,Roskes76,McKinstrie,Agrawal,Firth,Yu98,Rosenberg90,Haelterman94}.

Another important class of solutions of nonlinear equations are
solitary solutions ('solitons'). They may also exhibit MI if they
are localized in some dimensions but extended in one or more
others. MIs of the envelope solitons of the single NLS and of the
KdV solitons were pioneered, respectively, by Zakharov and
Rubenchik \cite{Zakharov74a} and  by Kadomtsev and Petviashvily
\cite{KP}. Latter MI  was studied in a number of other theoretical
and experimental works. For reviews on MI of bright and dark
solitary waves see, respectively,
\cite{PhysRep,Rasmussen,Berge,Akhmedbook,Saffman} and
\cite{Saffman,KivsharPHYSREP}.

From a formal point of view the problem of the solitary wave MI
can be considered as continuation of the soliton spectrum at zero
modulational frequency $\Omega$ into the region $\Omega\ne 0$. An
important class of  discrete eigenmodes at $\Omega=0$ are the zero
eigenvalue (or neutral) modes. Presence of which is directly
linked to symmetries of the model equations. On a qualitative
level, similarities and differences between MI of solitons and CW
solutions can be understood on the basis of a comparison between
the corresponding neutral modes. For example, the 1D, bright
spatial soliton of NLS  is modulationally unstable in media with
either anomalous or normal GVD. In first case, the neutral mode
associated with the phase symmetry is excited ('neck' MI) and in
latter situation the translational mode associated with a shift
along the direction perpendicular to the wave propagation becomes
unstable ('snake' MI) \cite{Zakharov74a}. The phase  mode is
present as well for CW solution and this leads to MI for anomalous
GVD. However, the translational  mode of CW solution is null and
therefore CWs are stable for normal GVD.

Increasing  the number of  free parameters can lead to more
complex scenarios of MI, because coexistence and competition
between different types of  instability are likely to happen. In a
recent Letter \cite{prl98} we considered GVD induced MI of spatial
solitons due to nondegenerate three-wave mixing. It was shown that
presence of an additional phase symmetry leads to the appearance
of a new branch of neck MI in media with normal GVD. It was found
that the novel instability strongly dominates the usual snake MI
throughout the region of soliton existence. Note, that because of
this dominance,  physical mechanisms responsible for the relative
strength of neck and snake instabilities remain to be understood.
Among  others, this last problem is addressed in the present work,
where we  study  MI of the bright solitary solutions in the
incoherently coupled NLS equations. The incoherent nature of the
coupling results in the presence of  two phase symmetries. In
spite of the similar symmetry properties, the features of MI
induced dynamics of solitons in the present model appear to be
more rich compare to the three-wave mixing case. In particular we
show that the relative strength of the nonlinear cross-coupling
governs the competition between neck and snake MIs in media with
normal GVD.

The rest of this paper is organized as follows.
In Section 2 we introduce model equations and discuss their
physical relevance. In Section 3 the problem
of MI of the solitary waves is formulated in general terms. MI of different
kinds of  solitary solutions and its physical interpretation in terms of
 polarization dynamics are detailed in Sections 4 and 5.
Discussion and summary of main results are presented in Sections 6 and 7.

\section{APPLICATIONS OF INCOHERENTLY COUPLED NLS EQUATIONS
TO  PROPAGATION OF ELECTROMAGNETIC WAVES}

The evolution  of  two suitably scaled slowly varying
incoherently coupled wave envelopes $E_1$ and $E_2$ in a weakly
nonlinear,  dispersive and diffractive medium is governed
by the following equations \cite{Berkhoer70}
\begin{eqnarray}
&& i\partial_z E_1+\alpha_1\vec\nabla_{\perp}^2E_1+\gamma_1\partial_t^2E_1+
(|E_1|^2+\beta |E_2|^2)E_1=0,\label{eq1}\\
\nonumber && i\partial_z E_2+\alpha_2\vec\nabla_{\perp}^2E_2
+\gamma_2\partial_t^2E_2+(|E_2|^2+\beta |E_1|^2)E_2=0,
\end{eqnarray}
where $\vec\nabla_{\perp}=\vec i\partial_x+\vec j\partial_y$.
Longitudinal $(z)$ and transverse $(x,y)$ coordinates are respectively measured in
units of a suitable diffraction length $l_{dif}$ and
of a characteristic transverse size of the envelope.
The coordinate $t$ is the retarded time scaled
to the parameter $T{\sqrt{l_{dif}/l_{dis}}}$, where $T$ is the
temporal duration of the envelope and
$l_{dis}$ is a characteristic GVD length.
Diffraction parameters $\alpha_{1,2}$ are positive while
GVD parameters $\gamma_{1,2}$  can have either sign.
Rescaling $x,y,t$ once more one can always choose $\alpha_1/\alpha_2$
and $|\gamma_1|/|\gamma_2|$ to be any convenient constants.
The parameter $\beta$ measures the relative strength of
cross-phase modulation compare to  self-phase modulation.
The nonlinearity was chosen to be self-focusing because below
we are interested in the dynamics of bright solitary waves.

Eqs. (1) describe a variety of physical situations
but we will focus here on their application to  propagation of
electro-magnetic (e/m) waves. Using a circular polarization basis
to describe propagation of quasimonochromatic e/m waves in isotropic dielectric
materials leads to Eqs. (1), where, in such a case,
 $E_1$ and $E_2$ are envelopes of the left- and right- polarized components \cite{Boyd}.
 The diffraction and GVD parameters
can be taken as $\alpha_{1,2}=0.5$, $\gamma_{1,2}=\gamma=\pm 0.5$, and
$\beta=1+\chi^{(3)}_{xyyx}/\chi^{(3)}_{xxyy}$
\cite{Boyd}, where $\chi^{(3)}_{ijkl}$ is the nonlinear
susceptibility tensor. For example,
$\beta=2$ for the nonresonant electronic nonlinearity and
$\beta=7$ for the nonlinearity due to molecular orientation \cite{Boyd}.
For e/m waves propagating in an isotropic plasma, $\beta$ strongly depends on the
ratio between the frequency of the e/m wave and the characteristic electron
plasma frequency, and it can have either sign \cite{McKinstrie}.
Eqs. (1) can also be applied to describe interaction of circularly polarised waves
in  waveguides filled with  linearly isotropic material, as, e.g.,
 $CS_2$ liquid  \cite{french}, in opposite situation linear coupling between
$E_1$ and $E_2$ should be incorporated in the model.

Counterpropagation of scalar waves in Kerr media
obeys  Eqs. (1) with $\beta$ determined by the wavelength-scale
refractive index gratings  written by the
interference pattern \cite{Firth}. The value of $\beta$ in this situation
is directly linked with diffusion which washes out the grating making
$1\le\beta\le 2$ ($\beta=2$ for zero diffusion).
Envelopes of incoherent copropagating waves in Kerr media
also obey Eqs. (1) with $\beta=2$ \cite{Agrawal}.
In these two situations the  group velocity difference
of  the wave envelopes, which is not explicitly written in Eqs. (1),
can be removed by a suitable phase shift.

The limiting case $\beta\to +\infty$ describes a situation with
zero self-modulation effects.  This approximates the so called cascading limit
of non-degenerate three-wave mixing
in the quadratically nonlinear media \cite{Tran}.
Therefore one can expect that for the large enough $\beta$ MI
of the solitary solutions of Eqs. (1) should be equivalent
to the MI of the three-wave solitons \cite{prl98} but
that it should well be different for the relatively small $\beta$.

Because the discussion of a wide range of $\beta$ values is more
realistic in the context of the interaction of the circularly
polarized waves, below we mainly use terminology which is
appropriate to this case.

\section{MODULATIONAL INSTABILITY OF SOLITONS.
GENERAL FORMULATION OF THE PROBLEM}

The primary target of the present paper is understanding of
the instabilities of the ground-state, i.e. nodeless, spatially localised
solutions of Eqs. (1) under the
action of the $t$-dependent perturbations. These solutions are well
known, see, e.g. \cite{eleon,malomedjosa,Ueda,Benney,Yang97,Yang,Malomed98} and refs.
therein. Here, we restrict ourselves
to the situation when the solitary waves are stable for $\partial_t=0$.
Therefore we choose $\beta>0$, because it ensures absence of the 'splitting'
instability \cite{Benney},  and $\vec\nabla_{\perp}=\vec i\partial_x$,
to avoid  collapse \cite{Rasmussen,Berge,Tsurumi}.

It is important for the following to summarize relevant symmetry
properties of Eqs. (1) with suppressed time derivatives
($\partial_t=0$). Invariance with respect to the two-parameter
gauge transformation
\begin{equation}
(E_1,E_2)\to(E_1e^{i\phi_1},E_2e^{i\phi_2}),
\label{eq2a}\end{equation}
leads to  conservation of the energies $Q_{1,2}=\int dx|E_{1,2}|^2$
or their equivalent combinations. There  are also invariances with
respect to  transverse translation and Galilean transformation,
\begin{eqnarray}
&& E_{1,2}(x)\to E_{1,2}(x+x_0), \label{eq2b}\\
&& E_{1,2}(x)\to E_{1,2}(x-vz)e^{iv(x- vz/2)}.\label{eq2c}
\end{eqnarray}
$\phi_1$, $\phi_2$, $x_0$ and $v$ are free parameters. Although we
use below the fact of the presence of the translational and
Galilean symmetries, we do not need explicit expressions for the
associated integrals of motion, which are linear momentum and
'center of mass', see, e.g., \cite{Akhmedbook}.

Symmetry property (\ref{eq2a}) indicates that the solitary
solutions can be presented in the form
\begin{eqnarray}
 E_{1,2}(x,z)=A_{1,2}(x)e^{i\kappa_{1,2}z}.
\label{eq2d}\end{eqnarray}
$A_{1,2}(x)$ are  real functions obeying the system of
ordinary differential equations
\begin{eqnarray}
 \frac{1}{2}\partial_x^2A_{1,2}=\kappa_{1,2}A_{1,2}-
(A_{1,2}^2+\beta A_{2,1}^2)A_{1,2}.\label{eq3}
\end{eqnarray}
Exponential localization of the solitons requires $\kappa_{1,2}>0$.
Actually one of these parameters can always be
scaled away, which means that fixing one of them and varying
the other in the whole region of the solitary wave existence one
will capture all possible situations. However,
for  convenience of analytical calculations it is better to
keep them both.

To study MI we seek solutions of Eq. (1)
in the form of  spatial solitons weakly modulated in time at frequency
$\Omega\ge 0$
\begin{eqnarray}
&& \nonumber E_{1,2}(x,z)=(A_{1,2}(x)+\\
&& (U_{1,2}(x,z)+iW_{1,2}(x,z))\cos\Omega t)
e^{i\kappa_{1,2}z+i\phi_{1,2}}.
\label{eq17a}\end{eqnarray}
Presenting solution of the linearized real system for the small perturbations
$U_{m},W_{m}$ in the form $U_m\sim u_m(x)e^{\lambda z}$ and
$W_m\sim w_m(x)e^{\lambda z}$
we obtain the following eigenvalue problem (EVP)
\begin{eqnarray}
&&(\hat{\cal L}_1+\gamma\Omega^2\hat I)\vec u=-\lambda\vec w, \label{eq100}\\
\nonumber &&(\hat{\cal L}_0+\gamma\Omega^2\hat I)\vec w=\lambda\vec u,
\end{eqnarray}
where $\vec u=(u_1,u_2)^T$, $\vec w=(w_1,w_2)^T$ and $\hat I$ is
the identity operator. $\hat{\cal L}_0$ and  $\hat{\cal L}_1$ are:
\begin{equation}
\hat{\cal L}_0=\left(\begin{array}{cc}
\hat{\cal D}_{1} &  0 \\
0 & \hat{\cal D}_{2}\end{array}\right),
\end{equation}
\begin{equation}
\hat{\cal L}_1=\left(\begin{array}{cc}
\hat{\cal B}_{1} &  -2\beta A_1A_2 \\
-2\beta A_1A_2 & \hat{\cal B}_{2}
\end{array}\right),\end{equation}
where $\hat{\cal D}_{1,2}=
-\frac{1}{2}\partial_x^2+\kappa_{1,2}-A_{1,2}^2-\beta A_{2,1}^2$ and
$\hat{\cal B}_{1,2}=
-\frac{1}{2}\partial_x^2+\kappa_{1,2}-3A_{1,2}^2-\beta A_{2,1}^2$.

By means of simple transformation one can reduce EVP (\ref{eq100})
to the following two EVPs for real and imaginary parts of the perturbations:
\begin{eqnarray}
&&(\hat{\cal L}_0+\gamma\Omega^2\hat I)(\hat{\cal L}_1+\gamma\Omega^2\hat
I)\vec u=-\lambda^2\vec u,
 \label{eq6a}\\
&&(\hat{\cal L}_1+\gamma\Omega^2\hat I)(\hat{\cal L}_0+\gamma\Omega^2\hat
I)\vec w=-\lambda^2\vec w. \label{eq6b}
\end{eqnarray}
EVPs  (\ref{eq6a}), (\ref{eq6b}) are adjoint to each other.
Therefore they have identical spectra and in case of instability
the imaginary and real parts of perturbations grow with the same rates.
To answer stability question it is thus enough to study only one of
the EVPs, and we concentrate below on the EVP (\ref{eq6b}).

Let suppose that $(\kappa_{1,2}+\gamma\Omega^2)=\xi_{1,2}\ge 0$. Then,
generally, $\lambda^2\in (-\infty,-\lambda_g^2)$ is a
continuous part of the spectrum with unbounded eigenfunctions,
where $\lambda_g=\min(\xi_1,\xi_2)$.
For particular cases when $\hat{\cal L}_1$ becomes a diagonal operator
the continuum splits into two independent bands,
$(-\infty,-\xi_{1,2}^2)$, corresponding to unboundness
of $w_1(x)$ and $w_2(x)$, respectively.
Eigenvalues which do not belong to the continuum
constitute the discrete part of the spectrum and have bounded eigenfunctions.
Stable eigenmodes with  eigenvalues obeying $-\lambda_g^2<\lambda^2<0$, called 'gap modes'.
Any other mode of the discrete spectrum, i.e. any eigenmode with $\lambda^2$ complex or
positive, renders the soliton unstable.
If  $\xi_1<0$ and/or $\xi_2 < 0$, the gap is closed, $\lambda_g=0$.

The procedure which we mainly follow to study stability
of different types of solitary solutions consists of  three basic steps.
First, using analytical and
numerical analysis we identify the discrete spectrum of  EVP
(\ref{eq6b}) for $\Omega=0$. Second, we develop perturbation theory
for the neutral eigenmodes in the low-frequency limit, $\Omega\ll 1$.
Third, we numerically build continuations of all
discrete eigenvalues into the region of finite positive $\Omega$.
We also allow for possible splitting of  discrete eigenvalues from
the edge of the continuum, but this was never actually observed.

\section{INSTABILITIES OF CIRCULARLY POLARIZED AND MANAKOV SOLITONS}

The single wave solitons of Eqs. (\ref{eq3}) corresponding to
right and left circular polarized e/m waves are
\begin{eqnarray}
&& A_1(x)={\sqrt{2\kappa_1}}sech{\sqrt{2\kappa_1}}x,~~A_2=0
\label{eq9},\\
&& A_1=0,~~A_2(x)={\sqrt{2\kappa_2}}sech{\sqrt{2\kappa_2}}x.
\label{eq10} \end{eqnarray}
 For these solutions, EVP (\ref{eq6b}), separates into two
independent scalar problems. Considering for example stability
of soliton with $A_1\ne 0$ we get:
\begin{eqnarray}
&& (\hat{\cal N}_1+\gamma\Omega^2)(\hat{\cal N}_0
+\gamma\Omega^2) w_1=-\lambda^2w_1, \label{eq10a}\\
&&\left(-\frac{1}{2}\partial_x^2
+\kappa_2+\gamma\Omega^2-\beta A_1^2\right)^2 w_2
=-\lambda^2 w_2\label{eq10b},
\end{eqnarray}
where $\hat{\cal N}_0=-\frac{1}{2}\partial_x^2+\kappa_1-A_1^2$,
$\hat{\cal N}_1=-\frac{1}{2}\partial_x^2+\kappa_1-3A_1^2$.

The operator on left-hand side of (\ref{eq10b}) has a nonnegative spectrum therefore
corresponding values of $\lambda_n^2$ ($n=0,1,2,3\dots$) must be
nonpositive, which means absence of the unstable eigenmodes.
In fact eigenvalue problem (\ref{eq10b}) can be
solved analytically, see, e.g. \cite{malomedjosa,Landau}. The eigenvalues are
\begin{equation}
\lambda_n^2=-\left(\kappa_2+\gamma\Omega^2-\frac{\kappa_1}{4}
\left[{\sqrt{8\beta+1}}-2n-1\right]^2\right)^2.
\label{eq11} \end{equation}

Eq. (\ref{eq10a}) is exactly an EVP arising in theory of MI of
solitons in single NLS equation and details of its analytical
and numerical investigations can be found in \cite{PhysRep,Berge,Zakharov74a}.
For the sake of completeness and comparison with
MI of the other types of solutions we  summarize the main results here.

The discrete spectrum of operator $\hat{\cal N}_1\hat{\cal N}_0$ consists of
two neutral eigenmodes which can be readily identified by applying
infinitesimal phase and Galilean  transforms to solitary wave solution.
These modes are $w_{1\phi_1}=A_1$ and $w_{1v}=xA_1$.
Infinitesimal translations and variations of $\kappa_1$
generate  two neutrally stable modes of the adjoint operator
$\hat{\cal N}_0\hat{\cal N}_1$
which are  $u_{1x}=\partial_xA_1$ and
$u_{1\kappa_1}=\partial_{\kappa_1}A_1$.
These modes obey the  identities:
$\hat{\cal N}_0 w_{1\phi_1}=0$,
$\hat{\cal N}_0 w_{1v}=-u_{1x}$,
$\hat{\cal N}_1 u_{1\kappa_1}=-w_{1\phi_1}$,
$\hat{\cal N}_1 u_{1x}=0$.

Following \cite{Zakharov74a} we assume $\Omega^2\ll 1$
and substitute the asymptotic expansions
\begin{equation}
w_1=( w_1^{(0)}+\Omega^2 w_1^{(1)}+\dots),\label{eq12a}
\end{equation}
 and
\begin{equation}
\lambda^2=\Omega^2
({\lambda^{(1)}}^2+\Omega^2{\lambda^{(2)}}^2+\dots)
\label{eq12b}\end{equation}
 into Eq. (\ref{eq10a}).
In the first two orders we have $\hat{\cal N}_1\hat{\cal N}_0
w^{(0)}=0$ and $\hat{\cal N}_1\hat{\cal N}_0
w^{(1)}=-{\lambda^{(1)}}^2 w^{(0)} -\gamma (\hat{\cal
N}_0+\hat{\cal N}_1) w^{(0)}$. Solution in leading order is $
w_1^{(0)}=C_{\phi_1}w_{1\phi_1}+C_v w_{1v}$, where $C_{\phi_1}$,
$C_v$ are constants. Orthogonality properties  $\langle
w_{1\phi_1},u_{1x}\rangle= \langle w_{1v},u_{1\kappa_1}\rangle=0$
(here and below $\langle \vec f,\vec g\rangle=\sum_m\int~dx
f_mg_m$) result in the independence of the branches produced by
the phase and Galilean neutral modes. Therefore $C_{\phi_1}$ and
$C_v$ are in fact independent constants. Solvability condition of
the 1st order problem gives
\begin{eqnarray}
&& {\lambda_{\phi_1}^{(1)}}^2= 2\gamma
\frac{Q_1}{\partial_{\kappa_1}Q_1}=4\gamma\kappa_1,\label{eq12}\\
 && {\lambda_{v}^{(1)}}^2= -2\gamma
\frac{\langle u_{1x},u_{1x}\rangle}{Q_1}= -\frac{4}{3}\gamma\kappa_1.
\label{eq13}\end{eqnarray}
Eqs. (\ref{eq12}), (\ref{eq13}) indicate onset of  instability
for either sign of $\gamma$. However, the character of the instability
depends on the sign of $\gamma$. For anomalous GVD $(\gamma>0)$
the spatially symmetric eigenmode becomes unstable leading to clustering
of the soliton stripe into filaments (neck MI).
For normal GVD  $(\gamma<0)$ an excitation
of the antisymmetric eigenmode leads to  spatial symmetry breaking and
bending of the solitary stripe along the temporal coordinate (snake MI).
The period of the modulations is approximately equal to $2\pi/\Omega_{max}$, where
$\Omega_{max}$ is the maximally unstable frequency.

\begin{figure}
\centerline { \epsfxsize=8cm  \epsffile{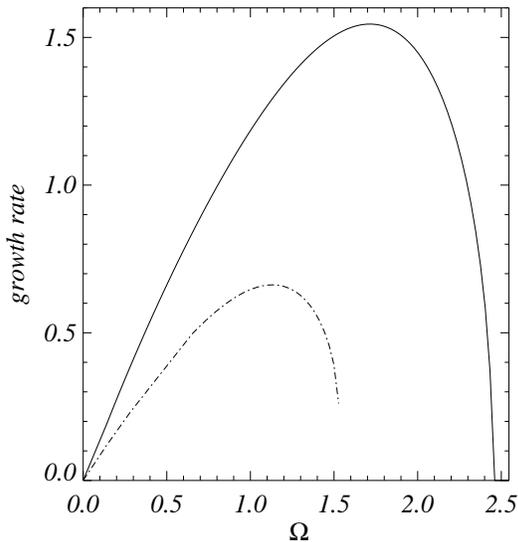}}
\caption{Instability growth rates of the circularly
polarized  soliton vs $\Omega$, $\kappa_1=1$.
Full (dot-dashed) line is for neck (snake) MI,
$\gamma=0.5$ ($\gamma=-0.5$).}
\label{fig1}\end{figure}

Typical dependencies of the MI growth rates vs $\Omega$ are
presented at Fig. \ref{fig1}. The neck instability  disappears at
$\Omega_{\phi_1}={\sqrt{3\kappa_1/\gamma}}$ where $w_1=0$ and
$u_1=sech^2{\sqrt{2\kappa_1}}x$. For $\Omega>\Omega_{\phi_1}$ the
corresponding eigenmode becomes a gap mode. Note that for
$\gamma>0$ the gap becomes wider with increasing of $\Omega$. The
snake instability disappearance  is very difficult to track
numerically because the corresponding eigenmode develops
oscillating tails and becomes weakly localized, so that a larger
number of the grid points is required. However, our numerical
analysis clearly indicates that the branch of snake MI does not
disappear stepwise at the point where the gap is closed,
$\Omega_{g}={\sqrt{-\kappa_1/\gamma}}$, as was suggested in Ref.
\cite{PhysRep}, but continues beyond this point and probably
reaches $\lambda^2=0$ at some larger $\Omega$.

The nonlinear stage of MI is also perfectly analogous to that in the
single NLS. Filaments, formed as the
result of the neck MI development, collapse to a
singularity during further propagation \cite{Rasmussen,Berge}.
The snake MI leads
to soliton spreading due to unbalanced action of the self-focusing
nonlinearity and normal GVD \cite{Berge}.
The second field $E_2$ is not affected by the discussed instabilities,
because of the incoherent nature of the coupling between $E_1$ and $E_2$.

In the special case $\beta=1$, $\alpha_1=\alpha_2$ and $\gamma_1=\gamma_2$
Eqs.(1) are invariant under the arbitrary rotations in  $(E_1,E_2)$ plane,
$E_{1,2}\to\cos\vartheta E_{1,2}\pm\sin\vartheta E_{2,1}$. This
 leads to a new parameterization of the ground
state solitons. These are usually called Manakov solitons \cite{Manakov}
 and they are  given by the
solutions of Eqs. (\ref{eq3}) with $\kappa_{1,2}=\kappa$:
\begin{equation}
A_1=\cos\theta A(x),~~
A_2=\sin\theta A(x).\label{eq20}
\end{equation}
Here the angle $\theta$ is a new free parameter
characterizing the polarization angle, and $A(x)={\sqrt{2\kappa}}
sech{\sqrt{2\kappa}}x$.
Because of the rotational invariance,
Manakov solitons with different polarizations are
equivalent and  their MI is indepent of the polarization angle.
Therefore one can always set $\theta=0$, and
then the corresponding EVP coincides with Eqs.
(\ref{eq10a}), (\ref{eq10b}).

\section{INSTABILITIES OF LINEARLY AND ELLIPTICALLY
POLARIZED SOLITONS ($\beta\ne 1$)}

\subsection{Soliton family and associated neutral modes}

To study solitons of an arbitrarily polarization for $\beta\ne 1$, i.e.
$A_1\ne 0$ and $A_2\ne 0$, it is more convenient to introduce absolute,
$\varphi=\frac{1}{2}(\phi_1+\phi_2)$, and relative,
$\psi=\frac{1}{2}(\phi_1-\phi_2)$,  phases.
The corresponding integrals of motion
are the total energy $Q=Q_1+Q_2$ and energy unbalance $Q_u=Q_1-Q_2$.
Associated soliton parameters are $\kappa=\frac{1}{2}(\kappa_1+\kappa_2)$
and $\delta=\frac{1}{2}(\kappa_1-\kappa_2)$.

For $\delta=0$ there is an obvious and well known analytical
solution of Eqs. (\ref{eq3})
\cite{Akhmedbook,malomedjosa,Ueda,Benney,Yang97,Yang,Malomed98}
\begin{equation}
A_{1,2}(x)=A(x)={\sqrt{\frac{2\kappa}{1+\beta}}}
sech{\sqrt{2\kappa}}x, \label{eq12c}\end{equation} corresponding
to a linearly polarized soliton. Using numerical solution of Eq.
(\ref{eq3}) one can verify that for $\beta\ne 1$ the exact
solution (\ref{eq12c}) belongs to the family of the solitary
solutions parameterized by $\kappa$ and $\delta$.

\begin{figure}
\centerline { \epsfxsize=8cm  \epsffile{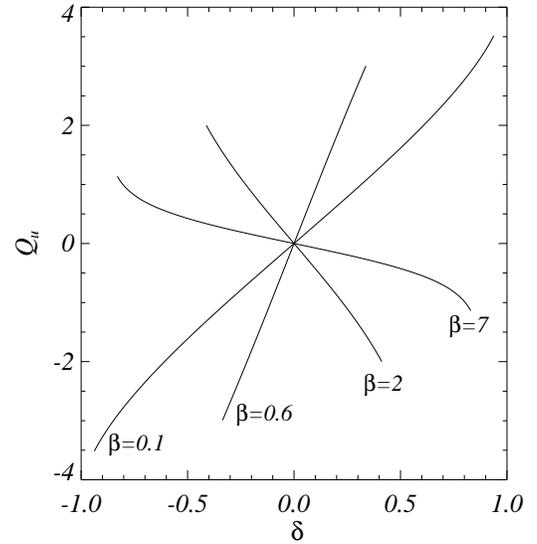}}
\caption{Energy unbalancing $Q_u$ vs $\delta$, $\kappa=1$.}
\label{fig2}\end{figure}

Using Eq. (\ref{eq11}) for $n=0$, $\Omega=0$
and its analog for the solution (\ref{eq10}) we conclude
that for  fixed values of $\kappa$ and $\beta$,
a family of ground state coupled solitary solutions
of Eqs. (\ref{eq3}) exists for $\delta\in (-\delta_c,\delta_c)$, where
\begin{equation}
\delta_{c}=\kappa\left|
\frac{1-4\beta+{\sqrt{1+8\beta}}}{3+4\beta-{\sqrt{1+8\beta}}}\right|.
\label{eq15} \end{equation} Analogs of Eqs. (\ref{eq15}) have been
derived before in a number of papers using different methods, see,
e.g. \cite{Akhmedbook,eleon,Yang}.  Expression under the modulus
in Eq. (\ref{eq15}) changes its sign from plus to minus once
$\beta$ changes from $\beta<1$ to $\beta>1$. It follows that for
$\beta<1$ the family of the elliptically polarized solitons splits
from the family $A_2=0$ ($A_1=0$) of the circularly polarized ones
at $\delta=\delta_c$ ($\delta=-\delta_c$) and this is {\em vice
versa} for $\beta>1$. Continuous variation of $\delta$ from
$-\delta_c$ to $\delta_c$ for fixed $\kappa$ and $\beta<1$
($\beta>1$) results in monotonic decay of $Q_2$ ($Q_1$) from its
maximal value $Q_{+}$ ($Q_{-}$) down to zero and in growth of
$Q_1$ ($Q_2$) from zero upto $Q_{+}$ ($Q_{-}$), where
$Q_{\pm}=2{\sqrt{2(\kappa\pm\delta_c)}}$. Therefore, we can make
an  important for the following conclusion, that for $\beta<1$
$\partial_{\delta}Q_u>0$ and for $\beta>1$
$\partial_{\delta}Q_u<0$. Numerically build dependencies of $Q_u$
vs $\delta$ for different values of $\beta$ are presented in Fig.
\ref{fig2}.

Consider now the main spectral properties of the elliptically
polarized solitons for $\Omega=0$ and  $\beta\ne 0,1$. Phase and
Galilean symmetries generate three neutral eigenmodes of the EVP
(\ref{eq6b}), they are $\vec w_{\varphi}=(A_1,A_2)^T$, $\vec
w_{\psi}=(A_1,-A_2)^T$, and $\vec w_{v}=x(A_1,A_2)^T$.
Infinitesimal variations of $\kappa$ and $\delta$, and
translational symmetry generate neutral modes of the adjoint
problem (\ref{eq6a}): $\vec
u_{\kappa}=\partial_{\kappa}(A_1,A_2)^T$, $\vec
u_{\delta}=\partial_{\delta}(A_1,A_2)^T$, and $\vec
u_{x}=\partial_x(A_1,A_2)^T$. These six modes obey the following
identities $\hat{\cal L}_0\vec w_{\varphi}=0$, $\hat{\cal L}_0\vec
w_{\psi}=0$, $\hat{\cal L}_0\vec w_{v}=-\vec u_x$, $\hat{\cal
L}_1\vec u_{\kappa}=-\vec w_{\varphi}$, $\hat{\cal L}_1\vec
u_{\delta}=-\vec w_{\psi}$, $\hat{\cal L}_1\vec u_{x}=0$.

For $\beta=0$ Eqs. (1) separate in two independent NLS equations.
The independence of the two fields results in additional
translational and Galilean symmetries characterizing freedom of
the relative transverse translation and motion of the two waves.
Therefore EVPs (\ref{eq6a}) and (\ref{eq6b}) have  additional
neutral modes $\vec u_{\delta x}=\partial_x(A_1,-A_2)^T$, $\vec
w_{\delta v}=x(A_1,-A_2)^T$. As  numerical solution for
$0<\beta<1$ shows the corresponding eigenvalue produces stable
branch of the discrete spectrum. For $|\beta|\ll 1$ approximate
expression for this eigenvalue can be readily found
\cite{Ueda,Benney}, $\lambda^2_{\delta v}=-64\beta/15$. Excitation
of the corresponding eigenmode results in position oscillations of
the soliton upon its propagation  \cite{Yang97}. When $\beta\to 1$
this eigenmode disappears into the continuum \cite{Yang97}.

\subsection{Asymptotic stability analysis $(\Omega^2\ll 1)$}

Now assuming that $\beta\gg\Omega^2$ we can use the
asymptotic techniques described in the previous section
to  continue zero-eigenvalue modes into the region  $\Omega^2\ll 1$.
Making substitutions
\begin{equation}
\vec w=( \vec w^{(0)}+\Omega^2\vec w^{(1)}+\dots)
\label{eq16}\end{equation}
and of Eq. (\ref{eq12b}) into (\ref{eq6b})
we get in the first two orders: $\hat{\cal L}_1\hat{\cal L}_0\vec w^{(0)}=0$ and
$\hat{\cal L}_1\hat{\cal L}_0 \vec w^{(1)}=-{\lambda^{(1)}}^2\vec w^{(0)}
-\gamma (\hat{\cal L}_0+\hat{\cal L}_1)\vec w^{(0)}$.
Solution in the leading order is
$\vec w^{(0)}=C_{\varphi}\vec w_{\varphi}
+C_{\delta}\vec w_{\delta}+C_v \vec w_{v}$.
As in the previous subsection one can show that
the branches produced by the two phase modes on the one hand
and by the Galilean  mode on the other are independent.
 The solvability condition of the first order problem
 for the Galilean mode gives
\begin{equation}
{\lambda_v^{(1)}}^2=
-2\gamma\frac{\langle \vec u_{x},\vec u_{x}\rangle }{Q},
\label{eq17}\end{equation}
which implies snake instability for $\gamma<0$.
When $\delta=0$, ${\lambda_v^{(1)}}^2=-4\gamma\kappa/3$,
cf. Eq. (\ref{eq13}).

For the two phase modes the solvability condition
results  in a quadratic equation for ${\lambda^{(1)}}^2$
\begin{equation}
a{\lambda^{(1)}}^4+b{\lambda^{(1)}}^2+c=0,
\label{eq18}\end{equation} where
$$4a=\partial_{\kappa}Q_u\partial_{\delta}Q
-\partial_{\kappa}Q\partial_{\delta}Q_u,$$ $$2b=\gamma
Q(\partial_{\kappa}Q+\partial_{\delta}Q_u)- \gamma
Q_u(\partial_{\kappa}Q_u+\partial_{\delta}Q),$$
$$c=-4\gamma^2Q_1Q_2.$$ Corresponding values of $C_{\varphi}$ and
$C_{\psi}$ are linked through the equality
$$\frac{C_{\varphi}}{C_{\psi}}= \frac{2\gamma
Q-{\lambda^{(1)}}^2\partial_{\delta}Q_u}
{{\lambda^{(1)}}^2\partial_{\delta}Q-2\gamma Q_u},$$ where
${\lambda^{(1)}}^2$ is the corresponding root of Eq. (\ref{eq18}).
In the general case expressions for the roots of Eq. (\ref{eq18})
can not be analyzed analytically, but it is already clear that
four symmetric neck type eigenmodes exist and either two or all of
them may be responsible for instability.

\begin{figure}
\centerline { \epsfxsize=8cm  \epsffile{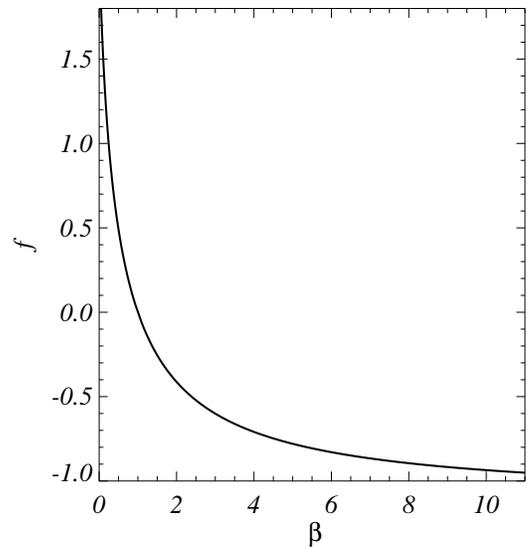}}
\caption{Function $f(\beta)$, see Eq. (\ref{eq19b}).}
\label{fig3}\end{figure}

If $\delta=0$ then $\langle\vec w_{\varphi},\vec u_{\delta}\rangle=
\langle\vec w_{\psi},\vec u_{\kappa}\rangle=0$
and therefore $\vec w_{\varphi}$ and  $\vec w_{\psi}$
eigenmodes produce independent branches of the discrete spectrum.
This results in  independence between $C_{\varphi}$ and $C_{\psi}$ and
  simplifies formulas for the associated eigenvalues:
\begin{eqnarray}
&& {\lambda^{(1)}_{\varphi}}^2=2\gamma\frac{Q}{\partial_{\kappa}Q}=
4\gamma\kappa,\label{eq19a}\\
&& {\lambda^{(1)}_{\psi}}^2=2\gamma\frac{Q}{\partial_{\delta}Q_u}=
2\gamma\kappa f(\beta).
\label{eq19b}\end{eqnarray}
Here $f(\beta)=(\int sech x~g(x,\beta)~dx)^{-1}$ and
function $g$ obeys to $(\partial_x^2-1+2\frac{3-\beta}{1+\beta}
sech^2x)g=sech x$. $f(\beta)$ changes its sign from
plus to minus when $\beta$ passes through  unity,  see Fig. \ref{fig3}.
Alternatively, Eq. (\ref{eq19b}) can be rewritten as
${\lambda^{(1)}_{\psi}}^2=2\gamma Q_1/\partial_{\delta}Q_1$,
cf. Eq. (4) in \cite{prl98}.

\begin{figure}
\centerline { \epsfxsize=8cm  \epsffile{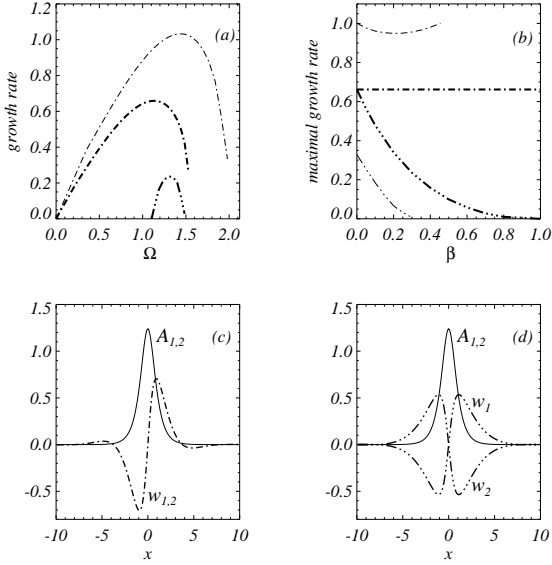}}
\caption{Instability growth rates, spatial profiles of the
solitary solutions and of the unstable eigenmodes for $\beta<1$,
$\gamma=-0.5$. Dash-dot (dash-dot-dot-dot) lines correspond to the
in-phase (anti-phase) snake MI. (a) Growth rates vs $\Omega$,
$\beta=0.3$. Thin (thick) lines correspond to $\kappa=1$,
$\delta=0$, $Q\simeq 4.04$, $Q_u=0$ ($\kappa=1.155$, $\delta=0.5$,
$Q\simeq 4.04$, $Q_u\simeq 2.18$). (b) Maximal growth rate vs
$\beta$. Thin (thick) lines correspond to $\delta=0$
$(\delta=0.5)$. (c) Components of the eigenmode corresponding to
the in-phase snake MI, $\beta=0.3$, $\delta=0$, $\Omega=1$. (d)
Components of the eigenmode corresponding to the anti-phase snake
MI, $\beta=0.3$, $\delta=0$, $\Omega=0.92$.}
\label{fig4}\end{figure}

The ${\lambda^{(1)}_{\varphi}}^2$ eigenvalue and associated
neutral mode $\vec w_{\varphi}$ are linked to the symmetry in the
absolute phase $\varphi$ and have their  analogies in the spectral
problem for  single wave solitons described in the previous
section, see Eq. (\ref{eq12}). The ${\lambda^{(1)}_{\psi}}^2$
eigenvalue and  neutral mode $\vec w_{\psi}$ are novel. They can
be directly attributed to the symmetry in the differential phase
$\psi$. This branch of the discrete spectrum generates instability
for normal GVD ($\gamma<0$) if $\beta>1$ and for anomalous GVD if
$\beta<1$, see Figs. \ref{fig4}, \ref{fig6}. Thus, the asymptotic
analysis indicates that for $\beta>1$, $\gamma<0$ neck and snake
instabilities coexist, and for $\beta<1$, $\gamma>0$ two different
types of  neck instability coexist. Numerical evaluation of the
roots of Eq. (\ref{eq18}) shows that the same conclusions hold
also for $\delta\ne 0$, throughout the whole existence region of
the family of  elliptically polarized solitons. Solving the EVP
(\ref{eq6b}) numerically, we find that in the low-frequency limit
the instability growth rates match those predicted by our
perturbation theory within of few percent up to $\Omega\simeq
0.5$. Numerical investigation (for more details see below) also
shows that apart from  the three instabilities discussed in
previous subsection, a fourth MI associated with continuation of
$\vec w_{\delta v}$ into the region of $\beta\ne 0$, $\Omega\ne 0$
also exists. Analytical treatment of this instability is also
possible, but it will not be pursued here, because corresponding
MI branch is never dominant.

Let us first discuss in general terms the physical meaning of all
the different types of the instabilities in the simple situation
with zero imbalancing ($\delta=0$), and only then we will proceed
with  details of the numerical analysis.

\subsection{Instability induced polarization dynamics}

All eigenmodes of the EVP (\ref{eq6b}) are two component vectors,
whose first and second components are responsible for the spatial
form of modulations of the fields $E_1$ and $E_2$, respectively.
The eigenmodes $\vec w_{\varphi}$ and $\vec w_v$ corresponding to
the variations of the absolute phase $\varphi$ and of the absolute
velocity $v$ of the coupled solitons have first and second
components which are in phase for any value of $x$. This property
holds  throughout the whole region of  existence of the associated
branches of the discrete spectrum. Therefore an excitation of
these eigenmodes is not accompanied by the breaking of the
polarization  of the initial state. In contrast, the eigenmodes
linked with $\vec w_{\psi}$ and $\vec w_{\delta v}$ neutral modes,
or in other words with variations of the relative phase $\psi$ and
the relative velocity $\delta v$, have anti-phased first and
second components. Therefore their excitation does lead to
polarization symmetry breaking. In particular, one should expect
that destabilization of the eigenmode associated with the relative
phase $\psi$ will result in breaking of the linearly polarized
soliton stripe into a chain of circularly polarized clusters,
where neighboring clusters have opposite (left and right)
polarizations. The same conclusions are obviously valid for  the
real parts of the perturbations.

Note here, that by  direct substitution of the linearly polarized
solution Eq. (\ref{eq12c}) into the EVPs (\ref{eq6a}),
(\ref{eq6b}) one can easily show that the eigenvalues of the
eigenmodes with in-phase first and second components are
independent of $\beta$. This means physically that in-phase MI is
insensitive to the relative strength of self- and cross-phase
modulations.

\subsection{Numerical results for normal GVD $(\gamma<0)$}

We start description of our numerical results with discussion of
the normal GVD case. We found two snake instabilities for
$\beta<1$, see Fig. 4. One of them corresponds to the in-phase
snaking of both fields, see Fig. 4(c), and its growth rate in the
low frequency limit is given by the Eq. (\ref{eq17}). The other
one corresponds to the anti-phase snaking, see Figs. 4(d).
Examples of the growth rate dependencies vs $\Omega$ and details
of the anti-phase snaking appearance are presented in Fig. 4(a)
and Fig. 5, respectively. Dependencies of the maximal instability
growth rates on $\beta$ are presented in Fig. \ref{fig4}(b). We
found that  in-phase snaking dominates  anti-phase snaking for all
values of $\delta$ and $\beta$. For $\beta=1$ the anti-phase snake
mode disappears inside the continuum and does not appear again for
all $\beta>1$.

The dominant role of the in-phase snake instability means that
breaking of the polarization state imposed by the initial
conditions is unlikely to happen upon propagation. Introducing
imbalancing for a fixed total energy enhances this dominance, see
Fig. 4(a). {\em Thus, when $\beta<1$, the linearly polarized
solitons are more stable than any other state of polarization.}

The anti-phase neck MI associated with the relative phase $\psi$
appears for $\beta>1$, see Eq. (\ref{eq19b}) and Figs. 6(b). The
in-phase snake instability obviously also exists, see Eq.
(\ref{eq17}) and Figs. 6(c),(d). The in-phase snake MI dominates
the anti-phase neck for $1<\beta<\beta_{sn}$ and  vice versa for
$\beta>\beta_{sn}$, see Fig. 7, where the cross-over value
$\beta_{sn}$ depends weakly on $\delta$. This fact can also be
seen from the comparison of the perturbative results for
$\delta=0$. According to the  Eqs. (\ref{eq19a}), (\ref{eq19b})
and Fig. (\ref{fig7}) the neck instability dominates the snake in
the low frequency limit starting from $\beta\simeq 3.55$.
Numerical stability analysis gives that $\beta_{sn}\simeq 3.47$ at
$\Omega=\Omega_{max}$ for $\delta=0$. Introducing  imbalancing
always leads to the suppression of the both instabilities, see
Figs. 6(a),(c) and Fig. 7. Therefore {\em the circular polarized
soliton is most stable for a given energy.}

In analogy with MI of  circularly polarized solitons for normal GVD,
the neck and snake unstable eigenmodes become weakly confined and
develop oscillating tails as $\Omega$ increases beyond the point
where the gap is closed, $\lambda_g=0$.

To test our linear stability analysis and study the nonlinear
evolution we performed  a series of computer simulations of the
Eqs. (1) with initial conditions in the form of a soliton stripe
perturbed by  spatio-temporal white noise of order of few percent.
Typical simulation results are presented in Fig.
\ref{fig9},\ref{fig10},\ref{fig11}. For $\beta<\beta_{sn}$ we
observed in-phase snaking of the stripe along the temporal
dimension, see Fig. \ref{fig9}. For $\beta>\beta_{sn}$ the soliton
stripe breaks in such a way as to form the interleaved intensity
peaks of $E_1$ and $E_2$, see Fig. \ref{fig10}, as expected when
the out of phase neck MI is dominant. The spatio-temporal patterns
formed at the initial stage of MI finally spread because of the
unbalanced action of the normal GVD and self-focusing
nonlinearity. For $\beta\simeq\beta_{sn}$ we observed competition
between the neck and snake MIs, see Fig. \ref{fig11}. In Fig.
\ref{fig11} (b$_1$), (b$_2$) one can clearly see that at the
intermediate stage of MI the typical in-phase snake pattern is
superimposed on the anti-phase neck pattern.

Thus, we conclude, that in the media with normal GVD spatial
soliton stripes perfectly develop snake MI without polarization symmetry
breaking if $\beta\in(0,\beta_{sn})$ and neck MI with polarization
symmetry breaking if $\beta\in(\beta_{sn},+\infty)$.

\subsection{Numerical results for anomalous GVD $(\gamma>0)$}

There are two neck MIs in this case for $\beta<1$, see Fig.
\ref{fig12}. One of them is associated with the absolute phase
$\varphi$ and corresponds to the in-phase neck MI. The other one
is associated with the relative phase $\psi$ and corresponds to
the anti-phase neck MI. The in-phase MI dominates the anti-phase
one for any value of $\delta$ and $\beta$, which means
conservation of the polarization state imposed by the initial
conditions. Nonzero imbalancing for fixed total energy leads to
the growth of the in-phase  MI and to the suppression of the
anti-phase one, see Fig. 11(a),(b). For $\beta>1$ only  in-phase
instability  exists, but now imbalancing leads to the suppression
of the instability, see Fig. 12. Presence of these instabilities
agrees with the predictions of low-frequency analysis, see Eqs.
(\ref{eq19a}), (\ref{eq19b}) and Figs. \ref{fig2},\ref{fig3}. A
typical result of the numerical simulation of the neck instability
development is shown in Fig. \ref{fig14}. Note that the
numerically attainable propagation distance was limited by the
distance at which the most intense filaments formed at the initial
stage of MI collapse to singularities.

Cut-off frequencies, where the neck MIs disappear can be found analytically for
$\delta=0$. Growth rate of the in-phase MI becomes zero at
$\Omega={\sqrt{3\kappa/\gamma}}$ in full analogy with
single NLS equation, see section IV. The anti-phase
MI disappears at $\Omega^2=\kappa(D-B)/(2\gamma)$
having $\vec w=0$ and $u_{1,2}=(sech {\sqrt{2\kappa}}x)^{(B-1)/2}$,
here $D=(11-5\beta)/(1+\beta)$, $B={\sqrt{(25-7\beta)/(1+\beta)}}$.

Thus,  in the media with anomalous GVD spatial
soliton stripe always develops neck MI without polarization symmetry
breaking and  filamentary structure formed during this process collapses
upon propagation. Detailed study of  collapse in  coupled NLS
equations is outwith the scope of this paper. Some details on this issue can
be found in  \cite{Tsurumi}.

\section{DISCUSSION}

It is interesting to compare MI of solitons with results on MI of CWs
\cite{Berkhoer70}, which can be easily recovered from Eqs.
(\ref{eq3}), (\ref{eq6a}),   (\ref{eq6b})
 putting $\partial_x^2=0$.
For simplicity we again consider the case of the linear
polarization, $E_{1,2}={\sqrt{\kappa/(1+\beta)}}e^{i\kappa z}$.
Then corresponding eigenvalues are
$\lambda^2_{\varphi}=\gamma\Omega^2(2\kappa-\gamma\Omega^2)$ and
$\lambda^2_{\psi}=\gamma\Omega^2(2\kappa(1-\beta)/(1+\beta)-\gamma\Omega^2)$.
For normal GVD, $\lambda^2_{\psi}$ can be positive only for
$\beta>1$. For anomalous GVD, $\lambda^2_{\varphi}$ generates
instability for any $\beta$ and $\lambda^2_{\psi}$ only for
$\beta<1$. Thus, as one could expect, neck instabilities of
solitons related to the phase symmetries have  analogies for CWs.
Snake instabilities are obviously absent for CWs, which is the
main difference between the dynamics of spatially confined
solitons and infinitely extended CWs. Namely, in the case of
normal GVD, CWs are modulationally  stable for $\beta<1$ and
unstable for $\beta>1$ (demonstrating polarization symmetry
breaking). Solitons are snake unstable in this situation for any
$\beta$ and this instability does not involve changes in the
polarization state. However, starting from a critical value of
$\beta=\beta_{sn}$ the snake instability becomes suppressed by the
neck one, which is analogous to instability of CW. This
instability does lead to polarization symmetry breaking. In
particular, a linearly polarized soliton breaks, due to this
instability, into the chain of circularly polarized clusters.
Because  snake instability leads to spatial symmetry breaking and
neck MI does not, the change in MI of solitons at
$\beta=\beta_{sn}$ can be interpreted as {\em a transition from
spatial symmetry breaking to polarization symmetry breaking.}

In the limit situation $\beta\gg 1$ self-phase effects are
negligible compare to cross-phase ones and development of the
in-phase and anti-phase neck MIs can be qualitatively explained
using Fermat's principle. Due to MI development the effective
refractive index for $E_1$ and $E_2$ fields gets modulated through
the XPM mechanism with period $2\pi/\Omega_{max}$. This results in
temporal cross-defocusing of filaments in media with normal GVD
and in cross-focusing for anomalous GVD. Thus, in the case of
normal GVD, interleaved pattern of the intensity peaks of $E_1$
and $E_2$ fields should be preferable because it enables each
field to see a refractive index that increases to its peak, i.e.
one that is in accord with Fermat's principle. This is clearly
verified in Fig.9.  The same arguments lead to the conclusion that
a pattern with all intensity peaks coincident is preferable for
anomalous GVD.

Considering possibility of experimental observations of predicted phenomena,
we have to say that diffraction induced MI of soliton-like stripe,
which is formally equivalent to the case of anomalous GVD, is
probably easiest to observe. However, it is less interesting at the same
time because it is perfectly
analogous to MI of CWs  and it is not accompanied by any polarization effects.
More interesting dynamics is expected in media with normal GVD.
In fact, experimental observation of temporal splitting
induced by normal GVD of  spatially confined pulses
in a self-focusing medium was recently reported in
\cite{Ranka96}.
However, transverse and polarization effects, which, accordingly to
our results, should play an important role, were not
studied during this experiment. Numerical studies
\cite{Ranka96} presented
to support the experimental results were restricted by scalar
approximation and radial geometry.

The rescaled instability growth rate $\lambda$ as function of the
modulational frequency $\Omega$ can be related to physical units
using the  formulae: 
$$\lambda_{ph}=\frac{\lambda}{4kw^2},~
\Omega_{ph}^2=\frac{\gamma\Omega^2}{2kk^{\prime\prime}w^2}.$$
Here $\lambda_{ph}$ and $\Omega_{ph}$ are the instability growth
rate and modulational frequency in physical units, $k$ is the wave
vector, $w$ is the beam width, $k^{''}=\partial_{\omega}^2k$.
 For example for radiation at $1\mu m$
propagating  in an AlGaAs planar waveguide  $k^{''}\simeq
-10^{-23}s^2/m$ \cite{belanger97josab}
 and for typical soliton transverse size $w\simeq 50\mu m$ \cite{aitchison97josab}
we get $\lambda_{ph}\simeq \lambda/(5cm)$ and
$\Omega_{ph}^2\simeq\Omega^2/(10^{-25}s^2)$. For
experiments with fused silica at wavelength $830nm$, see second
from Refs. \cite{Ranka96}, $k^{''}\simeq -10^{-26}s^2/m$ and
$\Omega_{ph}^2\simeq\Omega^2/(10^{-28}s^2)$.

\section{SUMMARY}

We have analyzed and described dispersive MI of families of
nodeless spatial solitons in the system of the two incoherently
coupled NLS equations. Considering coupled soliton states, we have
established the existence of the four branches of instabilities,
which are linked to the symmetries in the absolute and relative
phases and in the absolute and relative motions of solitons. We
gave a physical interpretation of our results describing
GVD-induced polarization dynamics. In particularly, we found that
in media with normal GVD the MI induced spatial symmetry breaking
in the transverse plane changes to the polarization symmetry
breaking when the relative strength of the cross-phase modulation
exceeds a certain threshold value. In media with anomalous GVD, MI
results in breaking of spatial solitons into spatio-temporal
clusters which collapse upon further propagation. This is not
followed by either spatial or polarization symmetry breaking.

\begin{figure}
\centerline { \epsfxsize=8cm  \epsffile{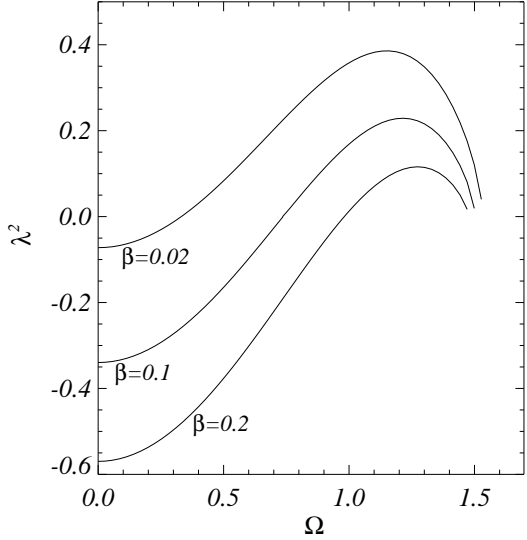}}
\caption{Eigenvalues corresponding to the anti-phase
snake MI vs $\Omega$ for several choices of $\beta$: $\kappa=1$,
$\delta=0$, $\gamma=-0.5$.}
\label{fig5}\end{figure}

\begin{figure}
\centerline { \epsfxsize=8cm  \epsffile{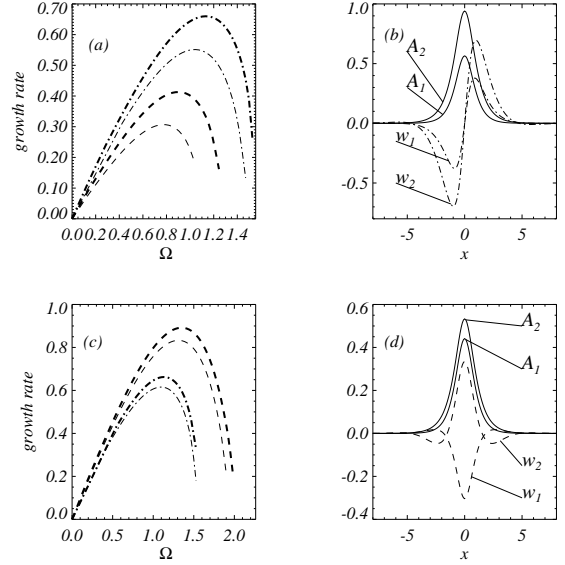}}
\caption{Instability growth rates, spatial profiles of the solitary solutions
 and  of the  unstable
eigenfunctions for $\beta>1$, $\gamma=-0.5$. Dash-dot (dash) lines
correspond to in-phase snake (anti-phase neck)  MIs. (a) Growth
rates vs $\Omega$, $\beta=2$. Thin (thick) lines correspond to
$\kappa=1$, $\delta=0$, $Q=1.75$, $Q_u=0$ ($\kappa=0.93$,
$\delta=0.2$, $Q=1.75$, $Q_u=-0.91$). (b) Components of the
eigenmode corresponding to in-phase snake MI, $\beta=2$,
$\delta=0.2$, $\Omega=0.8$. (c) Growth rates vs $\Omega$,
$\beta=7$. Thin (thick) lines correspond to $\kappa=1$,
$\delta=0$, $Q\simeq 0.68$, $Q_u=0$ ($\kappa=0.97$, $\delta=0.2$,
$Q\simeq 0.68$, $Q_u\simeq -0.16$). (d) Components of the
eigenmode corresponding to anti-phase neck MI, $\beta=7$,
$\delta=0.2$, $\Omega=1$.} \label{fig6}\end{figure}

\begin{figure}
\centerline { \epsfxsize=8cm  \epsffile{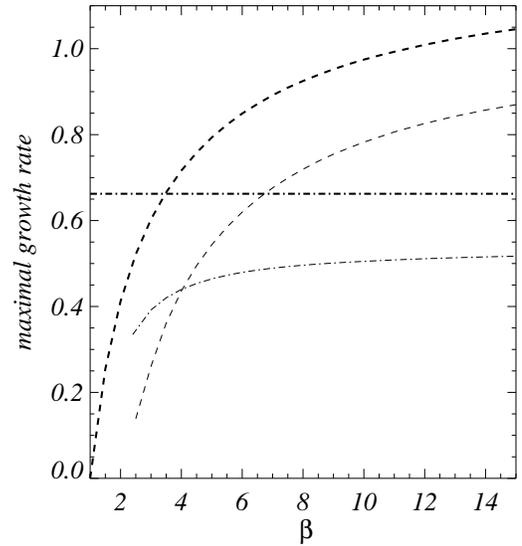}}
\caption{Maximal growth rates of the in-phase snake (dash-dot
line) and the anti-phase neck (dash line) MIs vs $\beta$ for
$\beta>1$, $\gamma=-0.5$. Thin (thick) lines correspond to
$\delta=0$ $(\delta=0.5)$. Cross-over occurs at $\beta_{sn}\simeq
3.47$ for $\delta=0$ and $\beta_{sn}\simeq 4.04$ for
$\delta=0.5$.} \label{fig7}\end{figure}

\begin{figure}
\centerline { \epsfxsize=8cm  \epsffile{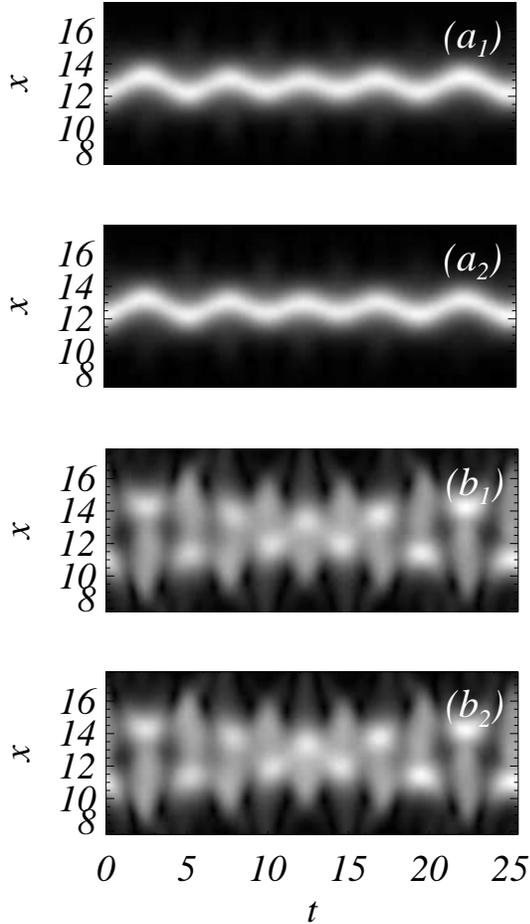}}
\caption{Development of the in-phase snake MI for $\beta=2$,
$\kappa=1$, $\delta=0$, $\gamma=-0.5$.
(a$_{1,2}$) $|E_{1,2}|$ for $z=12$; (b$_{1,2}$) $|E_{1,2}|$ for $z=14.7$.}
\label{fig9}\end{figure}

\begin{figure}
\centerline { \epsfxsize=8cm  \epsffile{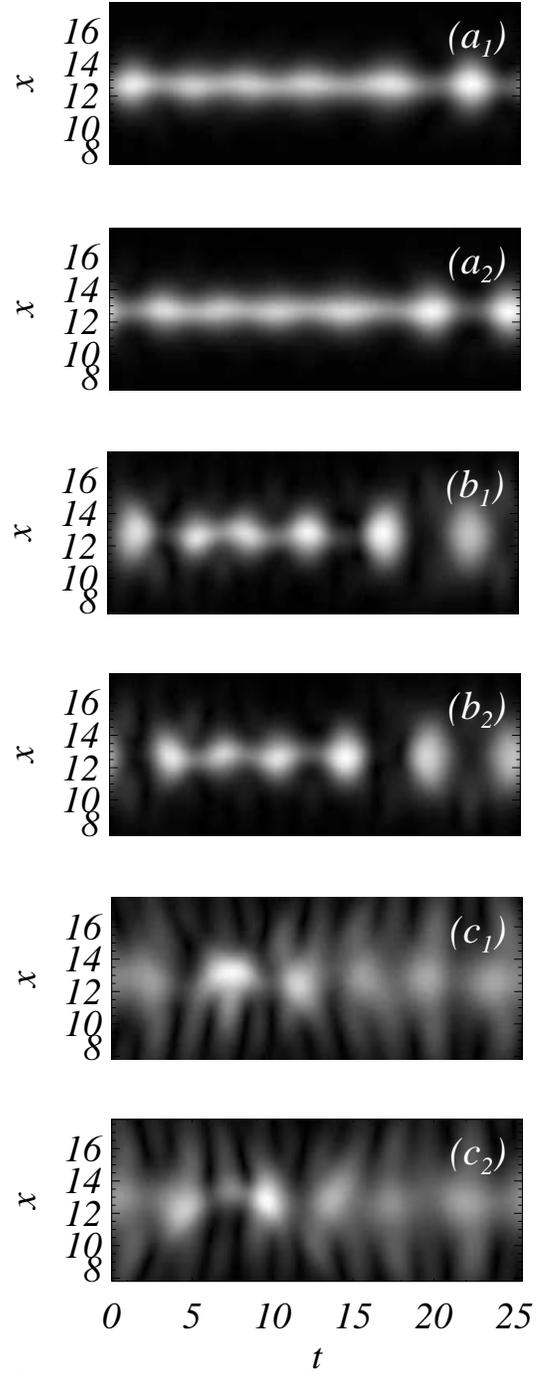}}
\caption{Development of the anti-phase neck MI for $\beta=7$,
$\kappa=1$, $\delta=0$, $\gamma=-0.5$.
(a$_{1,2}$) $|E_{1,2}|$ for $z=8.4$; (b$_{1,2}$) $|E_{1,2}|$ for $z=10.2$;
 (c$_{1,2}$) $|E_{1,2}|$ for $z=12.6$.}
\label{fig10}\end{figure}

\begin{figure}
\centerline { \epsfxsize=8cm  \epsffile{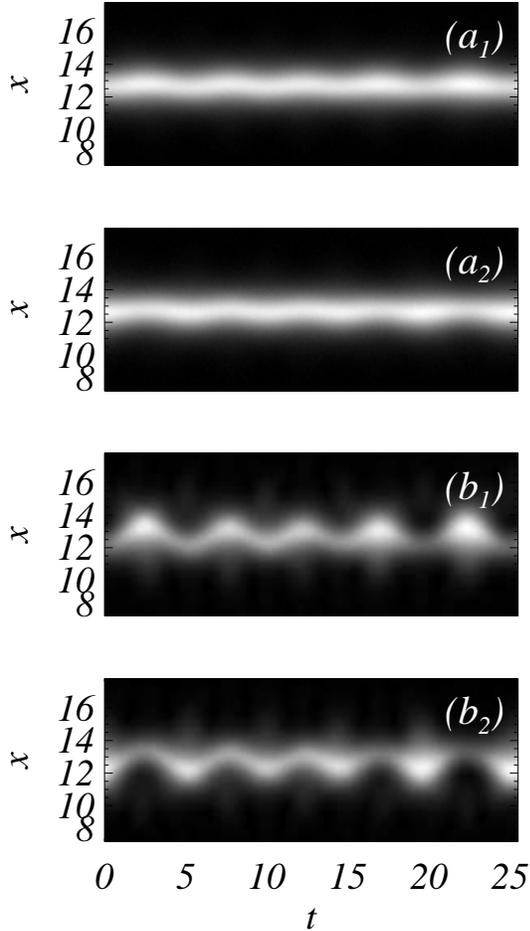}}
\caption{Competition between the in-phase snake and anti-phase neck MIs:
$\beta=3.47$, $\kappa=1$, $\delta=0$, $\gamma=-0.5$.
(a$_{1,2}$) $|E_{1,2}|$ for $z=9$; (b$_{1,2}$) $|E_{1,2}|$ for
$z=12$}
\label{fig11}\end{figure}

\begin{figure}
\centerline { \epsfxsize=8cm  \epsffile{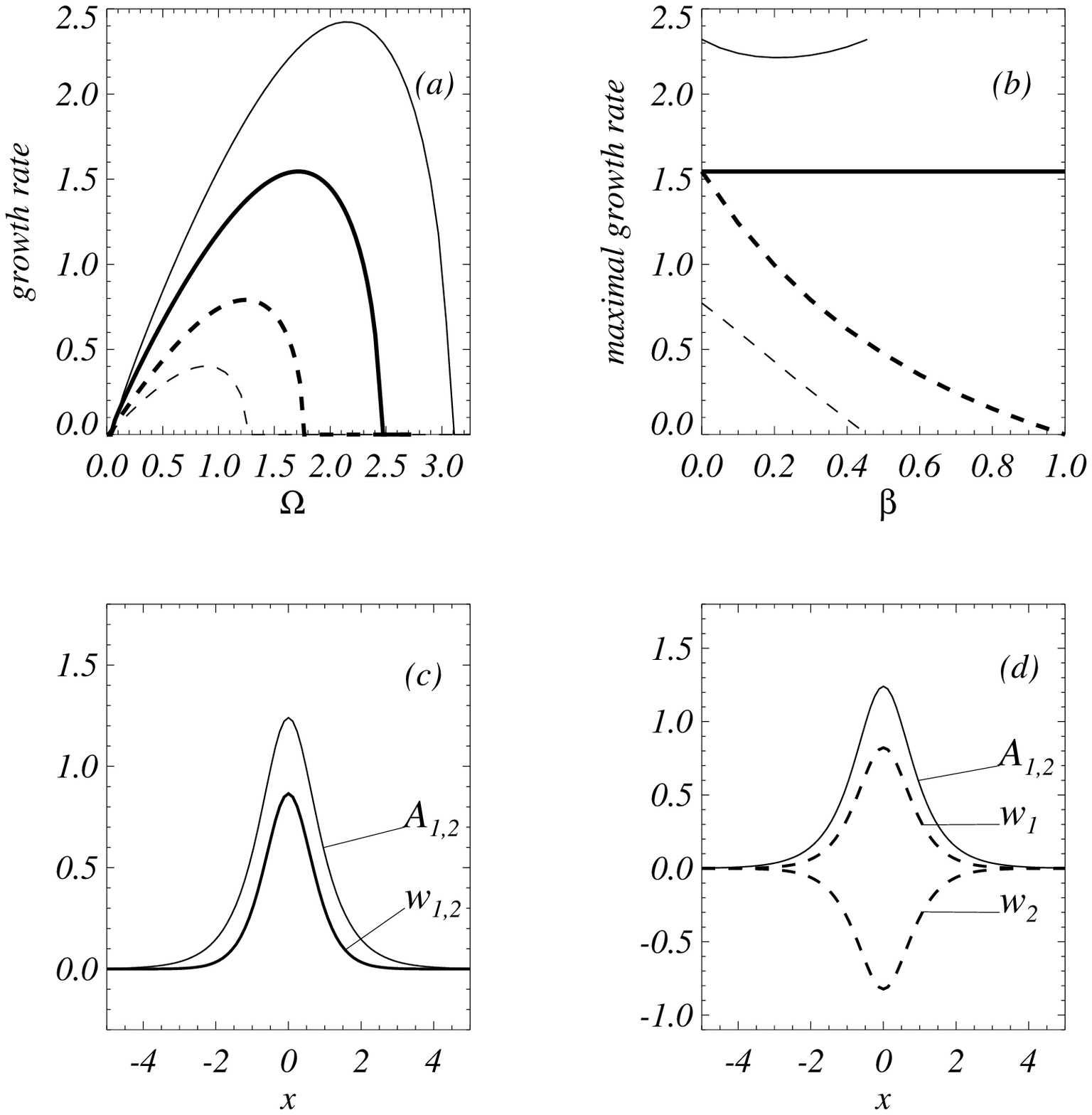}}
\caption{Instability growth rates, spatial profiles of the solitary solutions
 and  of  the unstable
eigenfunctions for $\beta<1$, $\gamma=0.5$. Full (dash) lines correspond
to in-phase (anti-phase) neck MIs.
(a) Growth rates vs $\Omega$, $\beta=0.3$. Thin (thick) lines correspond to
$\kappa=1$, $\delta=0$, $Q\simeq 4.04$, $Q_u=0$
($\kappa=1.155$, $\delta=0.5$, $Q\simeq 4.04$, $Q_u\simeq 2.18$).
(b) Maximal growth rate vs $\beta$. Thin (thick) lines
correspond to $\delta=0$ $(\delta=0.5)$. (c) Components of the eigenmode
corresponding to in-phase snake MI, $\beta=0.3$, $\delta=0$, $\Omega=1.5$.
(d) Componets of the eigenmode
corresponding to anti-phase snake MI, $\beta=0.3$, $\delta=0$,
$\Omega=0.9$.}
\label{fig12}\end{figure}

\begin{figure}
\centerline { \epsfxsize=8cm  \epsffile{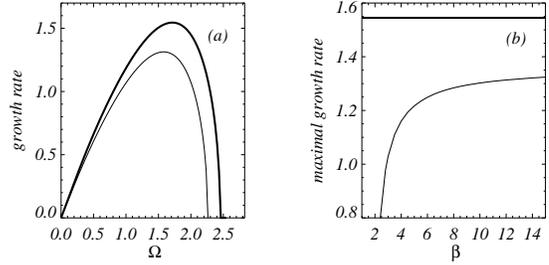}}
\caption{
In-phase MI growth rates for $\beta>1$, $\gamma=0.5$.
(a) Growth rates vs $\Omega$, $\beta=2$. Thin (thick) lines correspond to
$\kappa=1$, $\delta=0$, $Q\simeq 4.04$, $Q_u=0$
($\kappa=1.155$, $\delta=0.5$, $Q\simeq 4.04$, $Q_u\simeq 2.18$).
(b) Maximal growth rate vs $\beta$. Thin (thick) lines
correspond to $\delta=0$ $(\delta=0.5)$.  }
\label{fig13}\end{figure}

\begin{figure}
\centerline { \epsfxsize=8cm  \epsffile{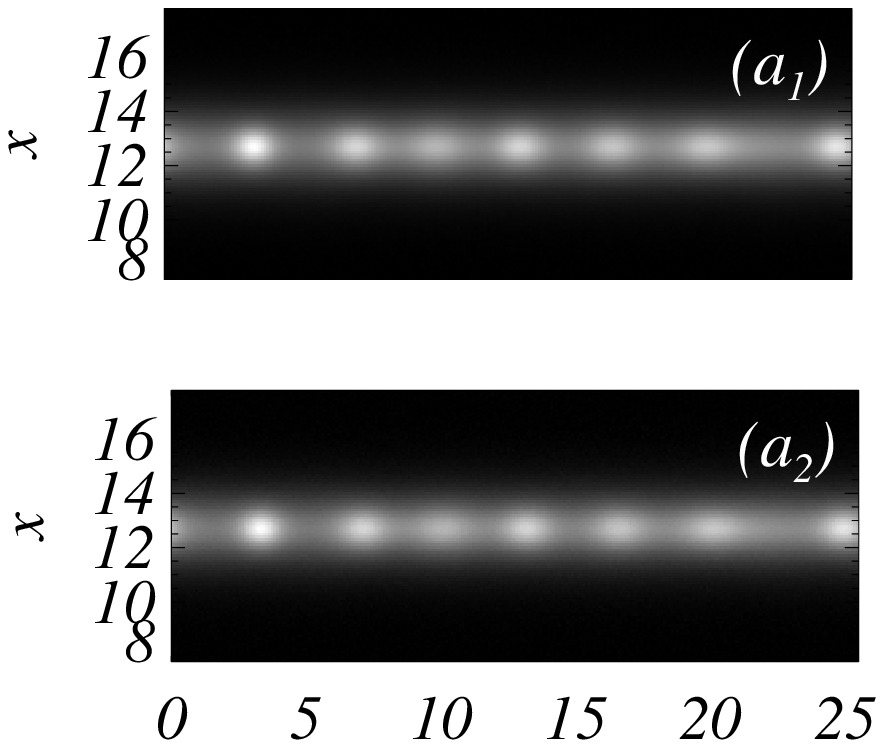}}
\caption{Development of the in-phase neck MI for $\beta=2$, $\kappa=1$,
$\delta=0$, $\gamma=0.5$.
(a$_{1,2}$) $|E_{1,2}|$ for $z=4.6$.
With further increasing of $z$ most intense filaments develop collapse.}
\label{fig14}\end{figure}


\begin{references}


\bibitem{Askaryan62}
G.A. Askar'yan, Zh. Eksp. Teor. Fiz. {\bf 42}, 1567 (1962)
[Sov. Phys. JETP {\bf 15}, 1088 (1962)];
T. Taniuti and H. Washimi, Phys. Rev. Lett. {\bf 21}, 209 (1968).

\bibitem{Bespalov66}
V.I. Bespalov and V.I. Talanov, Pis'ma Zh. Eksp. Teor. Fiz.
{\bf 3}, 471 (1966) [JETP Lett. {\bf 3}, 307 (1966)];
L.A. Ostrovskii, Zh. Eksp. Teor. Fiz. {\bf 51}, 1189 (1966)
[Sov. Phys. JETP {\bf 24}, 797 (1967)];
V.I. Karpman, Pis'ma Zh. Eksp. Teor. Fiz. {\bf 6}, 759 (1967)
[JETP Lett. {\bf 6}, 227 (1967)].

\bibitem{Benjamin67}
T.B. Benjamin and J.E. Feir, J. Fluid Mech. {\bf 27}, 417 (1967).

\bibitem{Pilipetskii65}
N.F. Pilipetskii and A.R. Rustamov, Pis'ma Zh. Eksp. Teor. Fiz.
{\bf 2}, 88 (1965) [JETP Lett. {\bf 2}, 55 (1965)].

\bibitem{comment1}  A comprehensive
review of  MI in hamiltonian systems is beyond our present scope,
therefore our list of references is restricted  to original works
\cite{Askaryan62,Bespalov66,Benjamin67,Pilipetskii65,KP}, some key
reviews  \cite{PhysRep,Rasmussen,Berge,Akhmedbook,KivsharPHYSREP}
and to papers which have direct relevance to the present work.


\bibitem{PhysRep}
E.A. Kuznetsov, A.M. Rubenchik, and V.E. Zakharov, Phys. Rep.  {\bf
142}, 103 (1986).

\bibitem{Rasmussen}
J.J. Rasmussen, K. Rypdal, Physica Scripta {\bf 33}, 481 (1986).

\bibitem{Berge}
L. Berge, Phys. Rep. {\bf 303}, 259 (1998).

\bibitem{Akhmedbook}
N.N. Akhmediev and A. Ankievich, {\em Solitons:
Nonlinear Pulses and Beams} (Chapman \& Hall, London, 1997) and refs. therein.

\bibitem{Saffman}
A.V. Mamaev, M. Saffman, D.Z. Anderson, and A.A. Zozulya,
Phys. Rev. A {\bf 54}, 870 (1996).

\bibitem{KivsharPHYSREP}
Y.S. Kivshar and B. Luther-Davies, Phys. Rep. {\bf 298}, 81 (1998).

\bibitem{book}
R.K. Dodd, J.C. Eilbeck, J.D. Gibbon, and H.C. Morris, {\em
Solitons and nonlinear wave equations} (Academic Press, London,
1984).

\bibitem{Berkhoer70}
A.L. Berkhoer and V.E. Zakharov, Zh. Eksp. Teor. Fiz. {\bf 58}, 903 (1970)
[Sov. Phys. JETP {\bf 31}, 486 (1970)].

\bibitem{Roskes76}
G.J. Roskes, Stud. Appl. Math. {\bf 55}, 231 (1976).

\bibitem{McKinstrie}
C.J. McKinstrie and R. Bigham, Phys. Fluids B {\bf 1}, 230 (1988);
G.G. Luther and C.J. McKinstrie, J. Opt. Soc. Am B {\bf 7}, 1125 (1990);
M. Yu, C.J. McKinstrie, and G.P. Agrawal, Phys. Rev. E {\bf 48} 2178 (1993).

\bibitem{Agrawal}
G.P. Agrawal, Phys. Rev. Lett. {\bf 59}, 880 (1987);
J. Opt. Soc. Am B {\bf 7}, 1072 (1990).

\bibitem{Firth}
W.J. Firth and C. Par\'e, Opt. Lett. {\bf 13}, 1096 (1988);
W.J. Firth, A. Fitzgerald, and C. Par\'e,
J. Opt. Soc. Am B {\bf 7}, 1087 (1990).

\bibitem{Yu98}
M. Yu, C.J. McKinstrie, and G.P. Agrawal, J. Opt. Soc. Am B {\bf
15}, 607 (1998).

\bibitem{Rosenberg90}
J.E. Rosenberg, Phys.  Rev. A {\bf 42}, R682 (1990).

\bibitem{Haelterman94}
M. Haelterman and A.P. Sheppard, Phys.  Rev. E {\bf 49}, 3389 (1994).


\bibitem{Zakharov74a}
V.E. Zakharov, Zh. Eksp. Teor. Fiz. {\bf 53}, 1735 (1967)
[Sov. Phys. JETP {\bf 26}, 994 (1968)];
V.E. Zakharov and A.M. Rubenchik, Zh. Eksp. Teor. Fiz. {\bf 65}, 997 (1973)
[Sov. Phys. JETP {\bf 38}, 494 (1974)].


\bibitem{KP}
B.B. Kadomtsev and V.I. Petviashvily, Dokl. Acad. Nauk SSSR {\bf 192}, 753 (1970)
[Sov. Phys. Dokl. {\bf 15}, 539 (1970)].


\bibitem{prl98}
D.V. Skryabin and W.J. Firth, Phys. Rev. Lett. {\bf 81}, 3379 (1998).

\bibitem{Boyd}
R.W.~Boyd, {\em Nonlinear Optics} (Academic Press, Boston, 1992).

\bibitem{french}
D. Wang, R. Barille, and G. Rivoire, J.~Opt. Soc. Am.~B {\bf 15}, 2731 (1998).

\bibitem{Tran}
H.T. Tran, Opt. Commun. {\bf 118}, 581 (1995).


\bibitem{eleon}
V.M. Eleonsky, V.G. Korolev, N.E. Kulagin, and L.P. Shilnikov, Zh.
Eksp. Teor. Fiz. {\bf 99}, 1113 (1991) [Sov. Phys. JETP {\bf 72},
619 (1991)]


\bibitem{malomedjosa}
B.A. Malomed, J. Opt. Soc. Am. B {\bf 9}, 2075 (1992).


\bibitem{Ueda}
T. Ueda and W.I. Kath, Phys. Rev. A {\bf 42}, 563 (1990).

\bibitem{Benney}
J. Yang and D.J. Benney, Stud. Appl. Math. {\bf 96}, 111 (1996).


\bibitem{Yang97}
J. Yang, Stud. Appl. Math. {\bf 98}, 61 (1997).

\bibitem{Yang}
J.K. Yang, Physica D {\bf 108}, 92 (1997).

\bibitem{Malomed98}
D.J. Kaup, B.A. Malomed and R.S. Tasgal, Phys. Rev. E {\bf 48}, 3049 (1993).
B.A. Malomed and R.S. Tasgal, Phys. Rev. E {\bf 58}, 2564 (1998).

\bibitem{Tsurumi}
C.J. McKinstrie and D.A. Russel, Phys. Rev. Lett. {\bf 61}, 2929 (1988);
A.A.~Afanas'ev, V.I. Kruglov, B.A. Samson, R. Jakyte, and V.M. Volkov,
J. Mod. Opt. {\bf 38}, 1189 (1991);
L.~Gagnon, J. Phys. A {\bf 25}, 2649 (1992);
T.~Tsurumi and M.~Wadati, J. Phys. Soc. Japan {\bf  67}, 93 (1998).


\bibitem{Landau}
L.D.~Landau and E.M.~Lifshitz, {\em Quantum Mechanics} (Pergamon Press, Oxford, 1965).

\bibitem{Manakov}
S.V. Manakov, Zh. Eksp. Teor. Fiz. {\bf 65}, 505 (1973)
[Sov. Phys. JETP {\bf 38}, 248 (1974)].

\bibitem{Ranka96}
J.K. Ranka, R.W. Schirmer, and A.L. Gaeta, Phys. Rev. Lett. {\bf 77}, 3783 (1996);
A.A. Zozulya, S.A. Diddams, A.G. Van Engen, and T.S. Clement,
 Phys. Rev. Lett. {\bf 82}, 1430 (1999).



\bibitem{belanger97josab}
N. Belanger, A. Villeneuve, and J.S. Aitchison, J. Opt. Soc. Am. B {\bf 14},
3003 (1997).

\bibitem{aitchison97josab}
J.S. Aitchison,  D.C. Hutchings, J.M. Arnold, J.U. Kang, G.I.
Stegeman, E. Ostrovskaya, and N. Akhmediev, J.~Opt. Soc. Am.~B
{\bf 14}, 3032 (1997).

\end{references}
\end{document}